\documentclass[twocolumn,showpacs,preprintnumbers,amsmath,amssymb,prb,floatfix]{revtex4}
\usepackage{graphicx}
\usepackage{dcolumn}
\usepackage{braket}
\usepackage{bm}

\begin{document}

\title{Electric dipole spin resonance with linear and cubic spin-orbit interaction}

\author{Yasuhiro Tokura}
\email{tokura.yasuhiro.ft@u.tsukuba.ac.jp}
\affiliation{
Faculty of Pure and Applied Sciences, University of Tsukuba,\\
Center for Artificial Intelligence Research, University of Tsukuba\\
}
\date{\today}

\begin{abstract}
We consider the electric dipole spin resonance (EDSR) with using the spin-orbit interaction (SOI)
in GaAs and Ge based quantum dots formed in a quantum well. 
We use Schrieffer-Wolff transformation and rotating frame to derive
the effective Hamiltonian of EDSR. 
We treat the couplings of the orbital motion with the environment with 
Gorini-Kossakowski-Sudarshan-Lindblad (GKSL) master equation.
We found that the cubic SOI makes the Rabi frequency non-linear with the
applied microwave amplitude.
At the same time, the fidelity of the spin manipulations becomes worse since 
the residual spin-orbital couplings induces spin relaxation.
\end{abstract}

\maketitle 

\section{Introduction}\label{sec:intro}
Coherent control of a single spin confined in a quantum dot (QD) 
by the microwave electric field, which is called as the electric dipole spin resonance (EDSR), 
is an active field of recent research. 
EDSR enables coherent spin manipulation by applying oscillating electric field, 
instead of applying oscillating magnetic field as traditionally done in the standard 
electron spin resonance (ESR).
One of the required mechanisms to couple the orbital motion to the spin degree 
of freedom is to utilize the slanting magnetic field\cite{Tokura}. 
This approach had enabled recent experimental studies, including 
fast Rabi oscillation\cite{Michel}, coupling spins with microwave photons\cite{Mi}, 
and fast QD electron spin manipulation\cite{Yoneda}. 
Another dominating mechanism uses intrinsic or extrinsic 
spin-orbit interactions (SOIs)\cite{Nowack}, 
which are particularly strong in hole spin systems\cite{Bulaev}.

Higher Rabi frequency is required for a given spin decoherence time 
to accomplish high-fidelity spin control for 
the fault-tolerant quantum computing.
In a standard analysis of EDSR, 
the Rabi frequency at resonant condition is proportional to the strength 
of SOI and the amplitude of the applied electric field. 
This relation holds in the lowest (linear) order in the linear field gradient or in 
linear-to-momentum SOI system, for example for electron spins 
with Rashba and linear Dresselhaus SOI in GaAs/AlGaAs heterostructures. 
In contrast, the behavior of the EDSR Rabi frequency for non-linear field gradient, 
non-parabolicity of the confinement, or cubic-to-momentum SOI are not yet well understood. 
The cubic SOI interaction exists for the Dresselhaus SOI and also Rashba 
SOI in Ge hole in a quantum well.
In this manuscript, we studied non-linear 
electric field amplitude dependence of the EDSR using cubic SOI.
We argue the effective Rabi Hamiltonian of these systems
and found that the spin and orbital degrees of freedom cannot be
decoupled even for the lowest order Schrieffer-Wolff transformation.
We then treat the dynamics of the orbital system coupled with the 
environment and is modeled with GKSL master equation.
Further, with restricting the orbitals to two levels, we derived effective
GKSL master equation for the spin with renormalized Rabi frequency.
We found the Rabi frequency for cubic SOI becomes super-linear with the applied
electric field amplitude and the spin relaxation rate increases quadratically
with the applied electric field.
In fact, there are several experimental reports on the Rabi frequency non-linearly
dependent on the applied microwave amplitude for example\cite{Yoneda, Takeda, Yoneda2, Noiri}.

This paper is organized as follows.
Section \ref{sec:model} explains the Hamiltonian of the model system and we 
apply Schrieffer-Wolff transformation to the Hamiltonian in Sec.~\ref{sec:swtr}.
Then in Sec.~\ref{sec:linear}, we start from the EDSR scheme which 
was first demonstrated in GaAs/AlGaAs system.
In the lowest order of SOI interaction and when we confine ourselves to a linear-type SOI only, 
the effective Hamiltonian for the spin is the ideal one where the Rabi frequency 
linearly depends on the applied electric field amplitude and the Rabi frequency 
depends on the direction of the field relative to the crystal orientation.
Then Sec.~\ref{sec:cubic} presents the main results of the EDSR with cubic SOI,
where the derived effective Hamiltonian contains residual spin-orbital couplings.
Section \ref{sec:dynamics} argues the dynamics of the orbital motion 
coupled to the environment with using GKSL master equation formalism, and then
derives effective GKSL master equation for the spin dynamics.
Finally, Sec.~\ref{sec:ge} discusses the EDSR of Rashba SOI in Ge hole system.
Section \ref{sec:conclusion} is the conclusions of this work and the
Appendix shows the details of the derivation of the effective GKSL master equation
for the spin using the adiabatic elimination approximation.

\section{model}\label{sec:model}
Let us consider one electron in a quantum dot (QD) subject to the uniform 
and constant magnetic field $\bm{B}$ and
time-dependent electric field $\bm{E}(t)$.
Without spin-orbit interaction, the magnetic field only affects the spin part 
of the electron wave function 
resulting in the Zeeman Hamiltonian
\begin{align}
\hat{\cal H}_{\mathrm{Z}}&=\frac{1}{2}g\mu_{\mathrm{B}} 
\bm{B}\cdot\hat{\bm{\sigma}}\equiv 
\frac{1}{2}E_{\mathrm{Z}}\bm{n}\cdot\hat{\bm{\sigma}},
\end{align}
where $g,\ \mu_{\mathrm{B}}$ are the Land\'{e} g-factor and Bohr magneton, 
respectively.
$\hat{\bm{\sigma}}\equiv (\hat{\sigma}_x,\hat{\sigma}_y,\hat{\sigma}_z)$ 
is the Pauli spin operator vector.
We have defined Zeeman energy $E_{\mathrm{Z}}\equiv g\mu_B |\bm{B}|$ 
and unit vector $\bm{n}$ pointing the direction of the magnetic field.

Similarly, the oscillating electric field only affects the orbital part 
of the electron wave function,
\begin{align}
\hat{\cal H}_{\mathrm{F}}(t)&=e\bm{E}(t)\cdot\hat{\bm{r}}=eE(t)\bm{n}'\cdot \hat{\bm{r}},
\end{align}
where $e$ is the unit charge, $\bm{n}'$ is the direction of the electric field 
and $\hat{\bm{r}}$ is the coordinate operator of the electron.

Static electron orbital Hamiltonian confined in a QD is
\begin{align}
\hat{\cal H}_{\mathrm{d}}&=\frac{\hat{\bm \pi}^2}{2m}+V(\hat{\bm{r}}),
\end{align}
where $\bm{\hat{\pi}}\equiv \hat{\bm p}+e\bm{A}(\hat{\bm{r}})$ 
is the generalized momentum with the vector potential 
$\bm{A}(\hat{\bm{r}})$ satisfying $\mbox{rot}\bm{A}=\bm{B}$.
The momentum operators satisfy commutation relation $[\hat{x}_\nu,\hat{p}_\mu]
=i\hbar \delta_{\nu\mu}$
and $[\hat{\pi}_x, \hat{\pi}_y]=ie\hbar B_z$.
$m$ is the electron effective mass.

$V(\hat{\bm{r}})$ is the static confinement potential of the QD. 
We restrict the discussion to the QD system made from two-dimensional electron system.
We set the $\hat{z}$ axis as normal to the plane and the confinement potential 
is factorized as
\begin{align}
V(\hat{\bm{r}})&=V_z(\hat{z})\cdot U(\hat{\bm{\rho}}),
\end{align}
where $V_z(\hat{z})$ is the confinement potential in the z-direction 
and $U(\hat{\bm{\rho}})$ is 
in-plane confinement potential where $\hat{\bm{\rho}}\equiv (\hat{x},\hat{y})$ 
is the two-dimensional coordinate operator.
We assume that the z-confinement is strong enough and 
we only need to consider the lowest 
quantum confined state. 
Therefore, in the following, we neglect any dynamics in the z-direction.
Moreover, for simplifying the discussion (except for the discussion on Ge hole system), 
we assume that the magnetic field 
and the (ac) electric field are  in-plane.
Therefore, the two-dimensional momentum operator is free 
from the vector potential: 
$\hat{\bm{\pi}}_\rho=\hat{\bm{p}}_\rho\equiv (\hat{p}_x,\hat{p}_y)$,
by choosing the gauge $\bm{A}=(0,0,B_x \hat{y}-B_y \hat{x})$.
Now the orbital Hamiltonian reduces to 
\begin{align}
\hat{\cal H}_{\mathrm{d2D}}&=\frac{\hat{\bm{p}}_\rho^2}{2m}+U(\hat{\bm{\rho}}).
\end{align}
In this note, we consider a harmonic potential $
U(\hat{\bm{\rho}})=m\omega_0^2(\hat{x}^2+\hat{y}^2)/2$,
where $\omega_0$ is the angular frequency of the potential.
Then, the orbital excitation energy is $E_{\mathrm{orb}}\equiv \hbar\omega_0$.

Finally, we define spin-orbit interaction (SOI) Hamiltonian. 
Assuming III-V semiconductor system, there are two types of 
spin-orbit interactions, which should be considered.
One is the Rashba-type SOI, depicted as
\begin{align}
\hat{\cal H}_{\mathrm{RSOI}}&=\alpha(\hat{p}_x\hat{\sigma}_y-\hat{p}_y\hat{\sigma}_x),
\end{align}
where the coefficient $\alpha$ is originated from the structure 
asymmetry in the z-direction and can be
tuned by the top gate voltage.
Another is the Dresselhaus-type SOI, which reads
\begin{align}
\hat{\cal H}_{\mathrm{DSOI}}&=\beta(-\hat{p}_x\hat{\sigma}_x
+\hat{p}_y\hat{\sigma}_y)+\gamma(\hat{p}_x\hat{p}_y^2\hat{\sigma}_x
-\hat{p}_y\hat{p}_x^2\hat{\sigma}_y)\notag\\
&\equiv \hat{\cal H}_{\mathrm{DSOI-L}}+\hat{\cal H}_{\mathrm{DSOI-C}},
\end{align}
where the first term with the coefficient $\beta$ is called linear Dresselhaus 
term and the second term with the coefficient $\gamma$ is 
called cubic Dresselhaus term.
In fact, $\beta$ and $\gamma$ is related by 
$\beta\equiv \gamma\left\langle \hat{p}_z^2\right\rangle$, 
where the average is taken over the
electron ground state wave function in the z-direction 
and hence can be tuned by deforming the wave function by the top gate voltage.
The parameter $\gamma$ is specific to the material, 
originating from the lack of the inversion symmetry in the crystal structure. 
Since both types of the SOI induces effective internal magnetic field 
normal to the momentum of the electron, 
we set $\bm{n}=\bm{n}'$, namely the static magnetic field and 
the ac electric field are parallel with each other, to accomplish the EDSR.
We define the total SOI Hamiltonian as a sum of linear and cubic terms
\begin{align}
\hat{\cal H}_{\mathrm{SOI}}&=\hat{\cal H}_{\mathrm{SOI-L}}+\hat{\cal H}_{\mathrm{SOI-C}},
\end{align}
where the linear term is
\begin{align}
\hat{\cal H}_{\mathrm{SOI-L}}&\equiv \hat{\cal H}_{\mathrm{RSOI}}
+\hat{\cal H}_{\mathrm{DSOI-L}}\notag\\
&=(-\beta\hat{p}_x-\alpha\hat{p}_y)
\hat{\sigma}_x+(\beta\hat{p}_y+\alpha\hat{p}_x)\hat{\sigma}_y\notag\\
&\equiv \hat{\xi}_x\hat{\sigma}_x+\hat{\xi}_y\hat{\sigma}_y
=\hat{\bm{\xi}}\cdot\hat{\bm{\sigma}}.
\end{align}
We define a vector operator $\hat{\bm{\xi}}$ with $\hat{\xi}_z=0$.
The total Hamiltonian is 
\begin{align}
\hat{\cal H}&=\hat{\cal H}_{\mathrm{d2D}}+\hat{\cal H}_{\mathrm{Z}}
+\hat{\cal H}_{\mathrm{SOI}}+\hat{\cal H}_{\mathrm{F}}(t).
\end{align}

\section{Schrieffer-Wolff transformation}\label{sec:swtr}
We introduce Schrieffer-Wolff transformation\cite{schrieffer-wolff} using an anti-Hermite 
operator $\hat{S}$, such that
\begin{align}
 e^{\hat{S}}\hat{\cal H}e^{-\hat{S}}
&=\hat{\cal H}+[\hat{S},\hat{\cal H}]+\frac{1}{2!}[\hat{S},[\hat{S},\hat{\cal H}]]+\cdots \notag\\
&=\hat{\cal H}_{\mathrm{d2D}}+\hat{\cal H}_{\mathrm{Z}}
+\hat{\cal H}_{\mathrm{F}}(t)+\hat{\cal H}_{\mathrm{SOI}}
+[\hat{S},\hat{\cal H}_{\mathrm{d2D}}+\hat{\cal H}_{\mathrm{Z}}]\notag\\
&+[\hat{S},\hat{\cal H}_{\mathrm{F}}(t)]+[\hat{S},\hat{\cal H}_{\mathrm{SOI}}]
+\frac{1}{2}[\hat{S},[\hat{S},\hat{\cal H}_{\mathrm{d2D}}
+\hat{\cal H}_{\mathrm{Z}}]]\notag\\
&+\frac{1}{2}[\hat{S},[\hat{S},\hat{\cal H}_{\mathrm{F}}(t)]]
+\frac{1}{2}\left[\hat{S}, \left[\hat{S}, \hat{\cal H}_{\mathrm{SOI}}\right]\right]+\cdots\notag\\
&=\hat{\cal H}_{\mathrm{SW}}+{\cal H}'+\mbox{higher orders}.
\end{align}
We assume that the operator $\hat{S}$ is the 
linear order in $\hat{\cal H}_{\mathrm{SOI}}$ and
satisfies following relation
\begin{align}\label{eq:eqofS}
[\hat{\cal H}_{\mathrm{d2D}}+\hat{\cal H}_{\mathrm{Z}}, \hat{S}]
&=\hat{\cal H}_{\mathrm{SOI}}.
\end{align}
Hence, the transformed Hamiltonian up to the first order in SOI is 
\begin{align}
\hat{\cal H}_{\mathrm{SW}}&=\hat{\cal H}_{\mathrm{d2D}}
+\hat{\cal H}_{\mathrm{Z}}+\hat{\cal H}_{\mathrm{F}}(t)+[\hat{S},\hat{\cal H}_{\mathrm{F}}(t)],
\end{align}
and the second order is
\begin{align}
\hat{\cal H}'&=\frac{1}{2}\left[\hat{S}, \hat{\cal H}_{\mathrm{SOI}}\right]
+\frac{1}{2}\left[\hat{S},\left[\hat{S}, \hat{\cal H}_{\mathrm{F}}(t)\right]\right].
\end{align}

In the following, we solve Eq.~(\ref{eq:eqofS}) for two separate factors,
\begin{align}
\hat{S}&\equiv \hat{S}_{\mathrm{L}}+\hat{S}_{\mathrm{C}},\\\label{eq:eqofSL}
\ [\hat{\cal H}_{\mathrm{d2D}}+\hat{\cal H}_{\mathrm{Z}},
\hat{S}_{\mathrm{L}}]&= \hat{\cal H}_{\mathrm{SOI-L}},\\
\ [\hat{\cal H}_{\mathrm{d2D}}+\hat{\cal H}_{\mathrm{Z}},
\hat{S}_{\mathrm{C}}]&=\hat{\cal H}_{\mathrm{DSOI-C}}.
\end{align}

\section{Effect of linear SOI}\label{sec:linear}
\subsection{Inverse of Liouville operator}
To make the analysis more transparent, we introduce the Liouville operator for any Hamiltonian 
$\hat{\cal H}_{\mathrm{A}}$ such that
\begin{align}
{\cal L}_{\mathrm{A}}\hat{B}&\equiv [\hat{\cal H}_{\mathrm{A}},\hat{B}],
\end{align}
where $\hat{B}$ is an arbitrary operator.
Then the equation for the linear SOI, Eq.(\ref{eq:eqofSL}), becomes
\begin{align}
({\cal L}_{\mathrm{d2D}}+{\cal L}_{\mathrm{Z}})\hat{S}_{\mathrm{L}}
&=\hat{\cal H}_{\mathrm{SOI-L}}.
\end{align}
Assuming there exists the inverse operator,
\begin{align}\label{eq:SL}
\hat{S}_{\mathrm{L}}&=({\cal L}_{\mathrm{d2D}}+{\cal L}_{\mathrm{Z}})^{-1}
\hat{\cal H}_{\mathrm{SOI-L}}\notag\\
&=\left\{{\cal L}_{\mathrm{d2D}}(1+{\cal L}_{\mathrm{d2D}}^{-1}
{\cal L}_{\mathrm{Z}})\right\}^{-1}\hat{\cal H}_{\mathrm{SOI-L}}\notag\\
&=(1+{\cal L}_{\mathrm{d2D}}^{-1}{\cal L}_{\mathrm{Z}})^{-1}
{\cal L}_{\mathrm{d2D}}^{-1}\hat{\cal H}_{\mathrm{SOI-L}}\notag\\
&\sim (1-{\cal L}_{\mathrm{d2D}}^{-1}{\cal L}_{\mathrm{Z}}
+{\cal L}_{\mathrm{d2D}}^{-1}{\cal L}_{\mathrm{Z}}
{\cal L}_{\mathrm{d2D}}^{-1}{\cal L}_{\mathrm{Z}}+\cdots)\notag\\
&\times {\cal L}_{\mathrm{d2D}}^{-1}\hat{\cal H}_{\mathrm{SOI-L}}\notag\\
&\equiv \hat{S}_{\mathrm{L}}^{(0)}+\hat{S}_{\mathrm{L}}^{(1)}+\hat{S}_{\mathrm{L}}^{(2)}+\cdots,
\end{align}
where we make an expansion assuming that the Zeeman energy 
is much smaller than the orbital excitation energy $E_{\mathrm{orb}}$, 
$E_{\mathrm{Z}}\ll E_{\mathrm{orb}}$.
Firstly, we focus our discussions upto the first order in $E_{\mathrm{Z}}$,
$\hat{S}_{\mathrm{L}}^{(1)}$.

We begin by studying the basic properties of Liouville operators,
\begin{align}
{\cal L}_{\mathrm{d2D}}\hat{x}&=[\frac{\hat{p}_x^2}{2m},
\hat{x}]=-\frac{i\hbar}{m}\hat{p}_x\equiv -iu\hat{p}_x,\\
{\cal L}_{\mathrm{d2D}}\hat{p}_x&=[\frac{m\omega_0^2}{2}\hat{x}^2,
\hat{p}_x]=i\hbar m\omega_0^2\hat{x}\equiv is\hat{x},
\end{align}
where we defined parameters $u\equiv \hbar/m$ and 
$s\equiv \hbar m\omega_0^2$.
Then in a matrix form
\begin{align}
{\cal L}_{\mathrm{d2D}}
\left(
  \begin{array}{c}
    \hat{x} \\
    \hat{p}_x
  \end{array}
\right)
&=
\left(
  \begin{array}{cc}
    0 & -iu  \\
    is & 0
  \end{array}
\right)
\left(
  \begin{array}{c}
    \hat{x} \\
    \hat{p}_x
  \end{array}
\right).
\end{align}
The inverse operator exists and is given by
\begin{align}
{\cal L}_{\mathrm{d2D}}^{-1}
\left(
  \begin{array}{c}
    \hat{x} \\
    \hat{p}_x
  \end{array}
\right)
&=
\left(
  \begin{array}{cc}
    0 & -\frac{i}{s}  \\
    \frac{i}{u} & 0
  \end{array}
\right)
\left(
  \begin{array}{c}
    \hat{x} \\
    \hat{p}_x
  \end{array}
\right).
\end{align}
Since this is a linear operator, we can apply twice
\begin{align}
\left({\cal L}_{\mathrm{d2D}}^{-1}\right)^2
\left(
  \begin{array}{c}
    \hat{x} \\
    \hat{p}_x
  \end{array}
\right)
&=
\left(
  \begin{array}{cc}
   \frac{1}{su} & 0  \\
   0 & \frac{1}{su}
  \end{array}
\right)
\left(
  \begin{array}{c}
    \hat{x} \\
    \hat{p}_x
  \end{array}
\right)
=\frac{1}{(\hbar\omega_0)^2}
\left(
  \begin{array}{c}
    \hat{x} \\
    \hat{p}_x
  \end{array}
\right).\notag\\
\end{align}
Similarly, we have
\begin{align}
{\cal L}_{\mathrm{d2D}}^{-1}
\left(
  \begin{array}{c}
    \hat{y} \\
    \hat{p}_y
  \end{array}
\right)
&=
\left(
  \begin{array}{cc}
    0 & -\frac{i}{s}  \\
    \frac{i}{u} & 0
  \end{array}
\right)
\left(
  \begin{array}{c}
    \hat{y} \\
    \hat{p}_y
  \end{array}
\right),
\end{align}
and 
\begin{align}
\left({\cal L}_{\mathrm{d2D}}^{-1}\right)^2
\left(
  \begin{array}{c}
    \hat{y} \\
    \hat{p}_y
  \end{array}
\right)
&=
\frac{1}{(\hbar\omega_0)^2}
\left(
  \begin{array}{c}
    \hat{y} \\
    \hat{p}_y
  \end{array}
\right).
\end{align}

We apply these to the linear SOI Hamiltonian 
\begin{align}
\hat{S}_{\mathrm{L}}^{(0)}&\equiv {\cal L}_{\mathrm{d2D}}^{-1}\hat{\cal H}_{\mathrm{SOI-L}}=
({\cal L}_{\mathrm{d2D}}^{-1}\hat{\bm{\xi}})\cdot
\hat{\bm{\sigma}}\equiv \hat{\bm{\zeta}}\cdot\hat{\bm{\sigma}}.
\end{align}
The operator vector $\hat{\bm{\zeta}}$ is evaluated as
\begin{align}
\hat{\zeta}_x&\equiv {\cal L}_{\mathrm{d2D}}^{-1}(-\beta\hat{p}_x
-\alpha\hat{p}_y)\notag\\
&=\frac{i}{u}(-\beta\hat{x}-\alpha\hat{y}),\\
\hat{\zeta}_y&=\frac{i}{u}(\beta\hat{y}+\alpha\hat{x}),
\end{align}
and $\hat{\zeta}_z=0$.
Then we apply the Zeeman part,
\begin{align}
{\cal L}_{\mathrm{Z}}({\cal L}_{\mathrm{d2D}}^{-1}\hat{\cal H}_{\mathrm{SOI-L}})
&=\left[\frac{1}{2}E_{\mathrm{Z}}\bm{n}\cdot\hat{\bm{\sigma}},
\hat{\bm{\zeta}}\cdot\hat{\bm{\sigma}}\right]\notag\\
&=\frac{1}{2}E_{\mathrm{Z}}n_i\hat{\zeta}_j[\hat{\sigma}_i,
\hat{\sigma}_j]\notag\\
&=iE_{\mathrm{Z}}n_i\hat{\zeta}_j\epsilon_{ijk}\hat{\sigma}_k\notag\\
&=iE_{\mathrm{Z}}(\bm{n}\times\hat{\bm{\zeta}})\cdot\hat{\bm{\sigma}},
\end{align}
where summation over repeated indices is assumed.
Finally,
\begin{align}
{\cal L}_{\mathrm{d2D}}^{-1}\left\{{\cal L}_{\mathrm{Z}}
({\cal L}_{\mathrm{d2D}}^{-1}\hat{\cal H}_{\mathrm{SOI-L}})\right\}&=
iE_{\mathrm{Z}}(\bm{n}\times {\cal L}_{\mathrm{d2D}}^{-1}
\hat{\bm{\zeta}})\cdot\hat{\bm{\sigma}}\notag\\
&=iE_{\mathrm{Z}}(\bm{n}\times ({\cal L}_{\mathrm{d2D}}^{-1})^2
\hat{\bm{\xi}})\cdot\hat{\bm{\sigma}}\notag\\
&=\frac{iE_{\mathrm{Z}}}{(\hbar\omega_0)^2}(\bm{n}
\times\hat{\bm{\xi}})\cdot\hat{\bm{\sigma}}\notag\\
&\equiv -\hat{S}_{\mathrm{L}}^{(1)}.
\end{align}
Therefore, to the first order in $E_{\mathrm{Z}}/\hbar\omega_0$, 
Eq.~(\ref{eq:SL}) becomes
\begin{align}
\hat{S}_{\mathrm{L}}^{(0+1)}&=\hat{\bm{\zeta}}\cdot \hat{\bm{\sigma}}
-\frac{iE_{\mathrm{Z}}}{(\hbar\omega_0)^2}(\bm{n}\times\hat{\bm{\xi}})
\cdot\hat{\bm{\sigma}}.
\end{align}
The factor related to $\hat{\bm{\xi}}$ is
\begin{align}
\bm{n}\times\hat{\bm{\xi}}&=(n_x\hat{\xi}_y-n_y\hat{\xi}_x)\bm{e}_z,
\end{align}
where $\bm{e}_z$ is the unit vector in the z-direction.
Therefore,
\begin{align}
\hat{S}_{\mathrm{L}}^{(0+1)}&=\frac{im}{\hbar}\left\{(-\beta\hat{x}-\alpha\hat{y})\hat{\sigma}_x
+(\beta\hat{y}+\alpha\hat{x})\hat{\sigma}_y\right\}\\
&-\frac{iE_{\mathrm{Z}}}{(\hbar\omega_0)^2}
\left\{n_x(\beta\hat{p}_y+\alpha\hat{p}_x)+n_y(\beta\hat{p}_x
+\alpha\hat{p}_y)\right\}\hat{\sigma}_z.\notag
\end{align}

\subsection{Higher order terms in Zeeman energy}
In this subsection, we further study higher order terms $\hat{S}_{\mathrm{L}}^{(\ell)}$
with $\ell\ge 2$.
First we evaluate
\begin{align}
&{\cal L}_{\mathrm{Z}}\left\{{\cal L}_{\mathrm{d2D}}^{-1}{\cal L}_{\mathrm{Z}}
{\cal L}_{\mathrm{d2D}}^{-1}\hat{\cal H}_{\mathrm{SOI-L}}\right\}
=\left[\frac{1}{2}E_{\mathrm{Z}}\bm{n}\cdot\hat{\bm{\sigma}},
\frac{iE_{\mathrm{Z}}}{(\hbar\omega_0)^2}(\bm{n}\times
\hat{\bm{\xi}})\cdot\hat{\bm{\sigma}}\right]\notag\\
&=\frac{iE_{\mathrm{Z}}^2}{2(\hbar\omega_0)^2}n_i(\bm{n}\times\hat{\bm{\xi}})_j
2i\epsilon_{ijk}\hat{\sigma}_k
=-\frac{E_{\mathrm{Z}}^2}{(\hbar\omega_0)^2}\bm{n}\times(\bm{n}\times\hat{\bm{\xi}})\cdot\
\hat{\bm{\sigma}},
\end{align}
where the vector is evaluated as
$\bm{n}\times(\bm{n}\times\hat{\bm{\xi}})=(n_x\hat{\xi}_y-n_y\hat{\xi}_x)(n_y, -n_x,0)$.
Explicitly, 
since ${\cal L}_{\mathrm{d2D}}^{-1}(n_x\hat{\xi}_y-n_y\hat{\xi}_x)
=n_x\hat{\zeta}_y-n_y\hat{\zeta}_x$,
we found that
\begin{align}
\hat{S}_{\mathrm{L}}^{(2)}&=-\frac{E_{\mathrm{Z}}^2}{(\hbar\omega_0)^2}
(n_x\hat{\zeta}_y-n_y\hat{\zeta}_x)(n_y\hat{\sigma}_x-n_x\hat{\sigma}_y)\notag\\
&=\frac{E_{\mathrm{Z}}^2}{(\hbar\omega_0)^2}(\bm{n}\times\hat{\bm{\zeta}})\cdot
(\bm{n}\times\hat{\bm{\sigma}}).
\end{align}

Next, 
\begin{align}
&{\cal L}_{\mathrm{Z}}\left\{\left({\cal L}_{\mathrm{d2D}}^{-1}{\cal L}_{\mathrm{Z}}\right)^2
{\cal L}_{\mathrm{d2D}}^{-1}\hat{\cal H}_{\mathrm{SOI-L}}\right\}\notag\\
&=\left[\frac{1}{2}E_{\mathrm{Z}}\bm{n}\cdot\hat{\bm{\sigma}},
\frac{E_{\mathrm{Z}}^2}{(\hbar\omega_0)^2}(\bm{n}\times\hat{\bm{\zeta}})\cdot
(\bm{n}\times\hat{\bm{\sigma}})\right]\notag\\
&=\frac{iE_{\mathrm{Z}}^3}{(\hbar\omega_0)^2}
(\bm{n}\times\hat{\bm{\zeta}})\cdot\hat{\bm{\sigma}},
\end{align}
where we used the relation
\begin{align}
\left[\bm{n}\cdot\hat{\bm{\sigma}},\bm{n}\times\hat{\bm{\sigma}}\right]&=
2i\left\{\hat{\bm{\sigma}}-(\bm{n}\cdot\hat{\bm{\sigma}})\bm{n}\right\}\notag
\end{align}
and note that $(\bm{n}\times\hat{\bm{\zeta}})\cdot\bm{n}=0$.
Then,
\begin{align}
\hat{S}_{\mathrm{L}}^{(3)}&=-{\cal L}_{\mathrm{d2D}}^{-1}
\frac{iE_{\mathrm{Z}}^3}{(\hbar\omega_0)^2}(\bm{n}\times\hat{\bm{\zeta}})
\cdot\hat{\bm{\sigma}}\notag\\
&=-\frac{iE_{\mathrm{Z}}^3}{(\hbar\omega_0)^4}(\bm{n}\times\hat{\bm{\xi}})
\cdot\hat{\bm{\sigma}}\notag\\
&=\left(\frac{E_{\mathrm{Z}}}{\hbar\omega_0}\right)^2\hat{S}_{\mathrm{L}}^{(1)}.
\end{align}
Hence, we can show that
\begin{align}
\hat{S}_{\mathrm{L}}^{(4)}
&={\cal L}_{\mathrm{d2D}}^{-1}{\cal L}_{\mathrm{Z}}(-\hat{S}_{\mathrm{L}}^{(3)})
=-\left(\frac{E_{\mathrm{Z}}}{\hbar\omega_0}\right)^2{\cal L}_{\mathrm{d2D}}^{-1}
{\cal L}_{\mathrm{Z}}\hat{S}_{\mathrm{L}}^{(1)}\notag\\
&=\left(\frac{E_{\mathrm{Z}}}{\hbar\omega_0}\right)^2
\hat{S}_{\mathrm{L}}^{(2)}.
\end{align}
Therefore, by iteration, we have
\begin{align}
\hat{S}_{\mathrm{L}}&=\hat{\bm{\zeta}}\cdot\hat{\bm{\sigma}}\notag\\
&-\frac{iE_{\mathrm{Z}}}{(\hbar\omega_0)^2}
\left\{1+\left(\frac{E_{\mathrm{Z}}}{\hbar\omega_0}\right)^2
+\left(\frac{E_{\mathrm{Z}}}{\hbar\omega_0}\right)^4+\cdots\right\}
(\bm{n}\times\hat{\bm{\xi}})\cdot\hat{\bm{\sigma}}\notag\\
&
+\left\{\left(\frac{E_{\mathrm{Z}}}{\hbar\omega_0}\right)^2
+\left(\frac{E_{\mathrm{Z}}}{\hbar\omega_0}\right)^4+\cdots\right\}
(\bm{n}\times\hat{\bm{\zeta}})\cdot
(\bm{n}\times\hat{\bm{\sigma}}).
\end{align}
Since we assume $|E_{\mathrm{Z}}|\ll \hbar\omega_0$, the infinite sum converges and 
we have
\begin{align}\label{eq:slfull}
\hat{S}_{\mathrm{L}}&=\hat{\bm{\zeta}}\cdot\hat{\bm{\sigma}}\notag\\
&-\frac{iE_{\mathrm{Z}}}{(\hbar\omega_0)^2-E_{\mathrm{Z}}^2}
(\bm{n}\times\hat{\bm{\xi}})\cdot\hat{\bm{\sigma}}\notag\\
&
+\frac{E_{\mathrm{Z}}^2}{(\hbar\omega_0)^2-E_{\mathrm{Z}}^2}
(\bm{n}\times\hat{\bm{\zeta}})\cdot
(\bm{n}\times\hat{\bm{\sigma}}).
\end{align}

\subsection{Effective Hamiltonian for linear SOI}
Now we evaluate the commutator 
$[\hat{S}_{\mathrm{L}},\hat{\cal H}_{\mathrm{F}}(t)]$.
Since $\hat{\cal H}_{\mathrm{F}}(t)$  and the first and the third term of 
$\hat{S}_{\mathrm{L}}$ in Eq.~(\ref{eq:slfull}) 
contains coordinate operators only, they commutes and hence
\begin{align}
&[\hat{S}_{\mathrm{L}},\hat{\cal H}_{\mathrm{F}}(t)]\notag\\
&=\Bigl[-\frac{iE_{\mathrm{Z}}}{(\hbar\omega_0)^2-E_{\mathrm{Z}}^2}
\left\{n_x(\beta\hat{p}_y+\alpha\hat{p}_x)
+n_y(\beta\hat{p}_x+\alpha\hat{p}_y)\right\}\hat{\sigma}_z,\notag\\
&eE(t)(n_x\hat{x}+n_y\hat{y})\Bigr]\notag\\
&=-\frac{\hbar E_{\mathrm{Z}}eE(t)}{(\hbar\omega_0)^2-E_{\mathrm{Z}}^2}
\left\{\alpha+2n_xn_y\beta\right\}\hat{\sigma}_z
\equiv -{\cal E}_{\mathrm{L}}(t)\hat{\sigma}_z,
\end{align}
where the effective oscillating Zeeman energy ${\cal E}_{\mathrm{L}}(t)$ is a c-number.
The total effective Hamiltonian is then reduced to
\begin{align}
\hat{\cal H}_{\mathrm{SW}}&=\hat{\cal H}_{\mathrm{d2D}}
+\hat{\cal H}_{\mathrm{F}}(t)+\hat{\cal H}_{\mathrm{Z}}
-{\cal E}_{\mathrm{L}}(t)\hat{\sigma}_z,
\end{align}
where the first two terms are only for the orbital part of 
the electron and the last two terms are only for the spin part, and
are in the form of standard electron spin resonance (ESR) Hamiltonian.
Let the applied microwave field as $E(t)\equiv E_0 \cos\omega t$.
Assuming $\alpha>0,\beta>0$, the factor ${\cal E}_{\mathrm{L}}(t)$ 
is the largest\cite{Nowack}  for $n_x=n_y=\pm \frac{1}{\sqrt{2}}$ and
\begin{align}
{\cal E}_{\mathrm{L}}(t)&=\frac{\hbar E_{\mathrm{Z}}eE_0}
{(\hbar\omega_0)^2-E_{\mathrm{Z}}^2}(\alpha+\beta)\cos\omega t,
\end{align}
and is the smallest for $n_x=-n_y=\pm \frac{1}{\sqrt{2}}$ and
\begin{align}
{\cal E}_{\mathrm{L}}(t)&=\frac{\hbar E_{\mathrm{Z}}eE_0}
{(\hbar\omega_0)^2-E_{\mathrm{Z}}^2}(\alpha-\beta)\cos\omega t.
\end{align}

\section{Effect of cubic term}\label{sec:cubic}
\subsection{Inverse of Liouville operator for cubic term}
Here we derive inverse of the Liouville operator for the cubic 
Dresselhaus SOI Hamiltonian, $\hat{\cal H}_{\mathrm{DSOI-C}}$.
With similar procedure as before, we start from the operation of 
${\cal L}_{\mathrm{d2D}}$ to $\hat{p}_x\hat{p}_y^2$:
\begin{align}
{\cal L}_{\mathrm{d2D}}(\hat{p}_x\hat{p}_y^2)&=is\hat{x}
\hat{p}_y^2+is\hat{p}_x\left\{\hat{y}, \hat{p}_y\right\},\notag\\
{\cal L}_{\mathrm{d2D}}(\hat{x}\hat{p}_y^2)&=-iu\hat{p}_x
\hat{p}_y^2+is\hat{x}\left\{\hat{y},\hat{p}_y\right\},\notag\\
{\cal L}_{\mathrm{d2D}}(\hat{p}_x\left\{\hat{y},\hat{p}_y\right\})
&=is\hat{x}\left\{\hat{y},\hat{p}_y\right\}+2is\hat{p}_x\hat{y}^2
-2iu\hat{p}_x\hat{p}_y^2,\notag\\
{\cal L}_{\mathrm{d2D}}(\hat{x}\left\{\hat{y},\hat{p}_y\right\})
&=-iu\hat{p}_x\left\{\hat{y},\hat{p}_y\right\}+2is\hat{x}\hat{y}^2
-2iu\hat{x}\hat{p}_y^2,\notag\\
{\cal L}_{\mathrm{d2D}}(\hat{p}_x\hat{y}^2)&=is\hat{x}\hat{y}^2
-iu\hat{p}_x\left\{\hat{y},\hat{p}_y\right\},\notag\\
{\cal L}_{\mathrm{d2D}}(\hat{x}\hat{y}^2)&=-iu\hat{p}_x\hat{y}^2
-iu\hat{x}\left\{\hat{y},\hat{p}_y\right\}.
\end{align}
where $\left\{\cdots\right\}$ is anti-commutator.
The result is summarized in a matrix form,
\begin{align}
&{\cal L}_{\mathrm{d2D}}
\left(
  \begin{array}{c}
    \hat{p}_x\hat{p}_y^2 \\
    \hat{x}\hat{p}_y^2 \\
    \hat{p}_x\left\{\hat{y},\hat{p}_y\right\} \\
    \hat{x}\left\{\hat{y},\hat{p}_y\right\} \\
    \hat{p}_x\hat{y}^2 \\
    \hat{x}\hat{y}^2
  \end{array}
\right)\\
&=
\left(
  \begin{array}{cccccc}
    0 & is & is & 0 & 0 & 0 \\
    -iu & 0 & 0 & is & 0 & 0 \\
    -2iu & 0 & 0 & is & 2is & 0 \\
    0 & -2iu & -iu & 0 & 0 & 2is \\
    0 & 0 & -iu & 0 & 0 & is \\
    0 & 0 & 0 & -iu & -iu & 0
  \end{array}
\right)
\left(
  \begin{array}{c}
    \hat{p}_x\hat{p}_y^2 \\
    \hat{x}\hat{p}_y^2 \\
    \hat{p}_x\left\{\hat{y},\hat{p}_y\right\} \\
    \hat{x}\left\{\hat{y},\hat{p}_y\right\} \\
    \hat{p}_x\hat{y}^2 \\
    \hat{x}\hat{y}^2
  \end{array}
\right),\notag
\end{align}
and the inverse exists, which is
\begin{align}
{\cal L}_{\mathrm{d2D}}^{-1}&=
\left(
  \begin{array}{cccccc}
    0 & \frac{i}{3u} & \frac{i}{3u} & 0 & 0 & \frac{2is}{3u^2} \\
    -\frac{i}{3s} & 0 & 0 & \frac{i}{3u} & -\frac{2i}{3u} & 0 \\
    -\frac{2i}{3s} & 0 & 0 & -\frac{i}{3u} & \frac{2i}{3u} & 0 \\
    0 & -\frac{2i}{3s} & \frac{i}{3s} & 0 & 0 & \frac{2i}{3u} \\
    0 & \frac{2i}{3s} & -\frac{i}{3s} & 0 & 0 & \frac{i}{3u} \\
    -\frac{2iu}{3s^2} & 0 & 0 & -\frac{i}{3s} & -\frac{i}{3s} & 0
  \end{array}
\right),
\end{align}
and its square is
\begin{align}
({\cal L}_{\mathrm{d2D}}^{-1})^2&=
\left(
  \begin{array}{cccccc}
    \frac{7}{9su} & 0 & 0 & \frac{2}{9u^2} & \frac{2}{9u^2} & 0 \\
    0 & \frac{7}{9su} & -\frac{2}{9su} & 0 & 0 & \frac{2}{9u^2} \\
    0 & -\frac{4}{9su} & \frac{5}{9su} & 0 & 0 & \frac{4}{9u^2} \\
    \frac{4}{9s^2} & 0 & 0 & \frac{5}{9su} & -\frac{4}{9su} & 0 \\
    \frac{2}{9s^2} & 0 & 0 & -\frac{2}{9su} & \frac{7}{9su} & 0 \\
    0 & \frac{2}{9s^2} & \frac{2}{9s^2} & 0 & 0 & \frac{7}{9su}
  \end{array}
\right).
\end{align}
We have the same inverse Liouville operators with just exchanging 
$(\hat{x}, \hat{p}_x)$ and $(\hat{y}, \hat{p}_y)$,
\begin{align}
\left(
  \begin{array}{c}
    \hat{p}_y\hat{p}_x^2 \\
    \hat{y}\hat{p}_x^2 \\
    \hat{p}_y\left\{\hat{x},\hat{p}_x\right\} \\
    \hat{y}\left\{\hat{x},\hat{p}_x\right\} \\
    \hat{p}_y\hat{x}^2 \\
    \hat{y}\hat{x}^2
  \end{array}
\right).
\end{align}
Now the operation of ${\cal L}_{\mathrm{d2D}}^{-1}
\hat{\cal H}_{\mathrm{DSOI-C}}$ is possible
\begin{align}
{\cal L}_{\mathrm{d2D}}^{-1}\hat{\cal H}_{\mathrm{DSOI-C}}
&=\gamma\left[{\cal L}_{\mathrm{d2D}}^{-1}(\hat{p}_x\hat{p}_y^2)\hat{\sigma}_x
-{\cal L}_{\mathrm{d2D}}^{-1}(\hat{p}_y\hat{p}_x^2)\hat{\sigma}_y\right]\notag\\
&=\gamma\Big[\left\{\frac{i}{3u}\hat{x}\hat{p}_y^2
+\frac{i}{3u}\hat{p}_x\left\{\hat{y},\hat{p}_y\right\}
+\frac{2is}{3u^2}\hat{x}\hat{y}^2\right\}\hat{\sigma}_x\notag\\
&-\left\{\frac{i}{3u}\hat{p}_x^2\hat{y}+\frac{i}{3u}
\left\{\hat{x},\hat{p}_x\right\}\hat{p}_y+\frac{2is}{3u^2}\hat{x}^2
\hat{y}\right\}\hat{\sigma}_y\Big]\notag\\
&\equiv \hat{\bm{\zeta}}_{\mathrm{C}}\cdot\hat{\bm{\sigma}},
\end{align}
and
\begin{align}
&({\cal L}_{\mathrm{d2D}}^{-1})^2\hat{\cal H}_{\mathrm{DSOI-C}}\notag\\
&=\gamma\Big[\left\{\frac{7}{9su}\hat{p}_x\hat{p}_y^2
+\frac{2}{9u^2}\hat{x}\left\{\hat{y},\hat{p}_y\right\}
+\frac{2}{9u^2}\hat{p}_x\hat{y}^2\right\}\hat{\sigma}_x\notag\\
&-\left\{\frac{7}{9su}\hat{p}_x^2\hat{p}_y+\frac{2}{9u^2}
\left\{\hat{x},\hat{p}_x\right\}\hat{y}+\frac{2}{9u^2}\hat{x}^2
\hat{p}_y\right\}\hat{\sigma}_y\Big]\notag\\
&\equiv \hat{\bm{\xi}}_{\mathrm{C}}\cdot\hat{\bm{\sigma}}.
\end{align}
As before,
\begin{align}
\hat{S}_{\mathrm{C}}&=\hat{\bm{\zeta}}_{\mathrm{C}}
\cdot\hat{\bm{\sigma}}-iE_{\mathrm{Z}}(\bm{n}
\times\hat{\bm{\xi}}_{\mathrm{C}})\cdot\hat{\bm{\sigma}}.
\end{align}

\subsection{Effective Hamiltonian for cubic SOI}
In order to evaluate the effective Hamiltonian, 
we need to calculate commutators of 
$\hat{\bm \xi}_{\mathrm{C}}, \hat{\bm \zeta}_{\mathrm{C}}$ 
with $\hat{x},\hat{y},\hat{p}_x,\hat{p}_y$.
Using the definition,
\begin{align}
\hat{\zeta}_{\mathrm{C}x}&=\frac{i\gamma}{3u}
\left\{\hat{x}\hat{p}_y^2+\hat{p}_x\left\{\hat{y},\hat{p}_y\right\}
+\frac{2s}{u}\hat{x}\hat{y}^2\right\},\\
\hat{\zeta}_{\mathrm{C}y}&=-\frac{i\gamma}{3u}
\left\{\hat{p}_x^2\hat{y}+\left\{\hat{x},\hat{p}_x\right\}
\hat{p}_y+\frac{2s}{u}\hat{x}^2\hat{y}\right\},\\
\hat{\xi}_{\mathrm{C}x}&=\frac{2\gamma}{9u^2}
\left\{\frac{7u}{2s}\hat{p}_x\hat{p}_y^2+\hat{x}\left\{\hat{y},
\hat{p}_y\right\}+\hat{p}_x\hat{y}^2\right\},\\
\hat{\xi}_{\mathrm{C}y}&=-\frac{2\gamma}{9u^2}
\left\{\frac{7u}{2s}\hat{p}_x^2\hat{p}_y+\left\{\hat{x},\hat{p}_x\right\}
\hat{y}+\hat{x}^2\hat{p}_y\right\},
\end{align}
we have
\begin{align}
\left[\hat{\zeta}_{\mathrm{C}x},\hat{x}\right]
&=\frac{\hbar\gamma}{3u}\left\{\hat{y},\hat{p}_y\right\},\\
\left[\hat{\zeta}_{\mathrm{C}y},\hat{x}\right]
&=-\frac{2\hbar\gamma}{3u}(\hat{p}_x\hat{y}+\hat{x}\hat{p}_y),\\
\left[\hat{\zeta}_{\mathrm{C}x},\hat{y}\right]
&=\frac{2\hbar\gamma}{3u}(\hat{x}\hat{p}_y+\hat{p}_x\hat{y}),\\
\left[\hat{\zeta}_{\mathrm{C}y},\hat{y}\right]
&=-\frac{\hbar\gamma}{3u}\left\{\hat{x},\hat{p}_x\right\},\\
\left[\hat{\xi}_{\mathrm{C}x},\hat{x}\right]
&=-\frac{2i\hbar\gamma}{9u^2}\left(\frac{7u}{2s}\hat{p}_y^2+\hat{y}^2\right),\\
\left[\hat{\xi}_{\mathrm{C}y},\hat{x}\right]
&=\frac{4i\hbar\gamma}{9u^2}\left(\frac{7u}{2s}\hat{p}_x\hat{p}_y+\hat{x}\hat{y}\right),\\
\left[\hat{\xi}_{\mathrm{C}x},\hat{y}\right]
&=-\frac{4i\hbar\gamma}{9u^2}\left(\frac{7u}{2s}\hat{p}_x\hat{p}_y+\hat{x}\hat{y}\right),\\
\left[\hat{\xi}_{\mathrm{C}y},\hat{y}\right]
&=\frac{2i\hbar\gamma}{9u^2}\left(\frac{7u}{2s}\hat{p}_x^2+\hat{x}^2\right).
\end{align}
Using above results, we have
\begin{align}
[\hat{S}_{\mathrm{C}},\hat{\cal H}_{\mathrm{F}}(t)]
&=[\hat{\zeta}_{\mathrm{C}x}\hat{\sigma}_x+\hat{\zeta}_{\mathrm{C}y}\hat{\sigma}_y
-iE_{\mathrm{Z}}(n_x\hat{\xi}_{\mathrm{C}y}
-n_y\hat{\xi}_{\mathrm{C}x})\hat{\sigma}_z,\notag\\
&eE(t)(n_x\hat{x}+n_y\hat{y})]\notag\\
&=eE(t)\frac{m\gamma}{3}\notag\\
&\times\Big[\left\{n_x(\hat{y}\hat{p}_y
+\hat{p}_y\hat{y})+2n_y(\hat{x}\hat{p}_y+\hat{p}_x\hat{y})\right\}\hat{\sigma}_x\notag\\
&-\left\{2n_x(\hat{p}_x\hat{y}
+\hat{x}\hat{p}_y)+n_y(\hat{x}\hat{p}_x+\hat{p}_x\hat{x})\right\}\hat{\sigma}_y\notag\\
&+\frac{2mE_{\mathrm{Z}}}{3\hbar}
\Big\{2\left(\frac{7\hat{p}_x\hat{p}_y}{2(m\omega_0)^2}+\hat{x}\hat{y}\right)\notag\\
&+n_xn_y\left(\frac{7(\hat{p}_x^2
+\hat{p}_y^2)}{2(m\omega_0)^2}+\hat{x}^2+\hat{y}^2\right)\Big\}\hat{\sigma}_z\Big].
\end{align}
Therefore, the orbital and spin parts are not decoupled 
in contradict with the conclusion of Ref.~\cite{Golovach}.

\subsection{Rotate the coordinate system and Bosonic representation}
In this section, we rotate the coordinate system such that
the $x$-axis parallel to the external magnetic field and oscillating field.
The elements in new coordinate system are with primes,
\begin{align}
x'=n_x x+n_y y,\ & \ y'=n_x y-n_y x,\notag\\
p_x'=n_x p_x+n_y p_y,\ & \ p_y'=n_x p_y-n_y p_x,\notag\\
\hat{\sigma}_x'=n_x\hat{\sigma}_x+n_y \hat{\sigma}_y, & \ 
\hat{\sigma}_y'=n_x \hat{\sigma}_y-n_y\hat{\sigma}_x'.
\end{align}
After some manipulations, we have
\begin{align}
&\left[\hat{S}_{\mathrm{C}}, \hat{\cal H}_{\mathrm{F}}(t)\right]
=eE(t)\frac{m\gamma}{3}\notag\\
\times&\Bigl(\left[(n_x^2-n_y^2)\left\{\hat{y}',\hat{p}_y'\right\}
+2n_xn_y\left(\hat{x}'\hat{p}_y'+\hat{y}'\hat{p}_x'\right)\right]\hat{\sigma}_x'\notag\\
&-\Bigl[3n_xn_y\left\{\hat{x}',\hat{p}_x'\right\}-n_xn_y\left\{\hat{y}',\hat{p}_y'\right\}\notag\\
&\hspace{2.5cm}
+2(n_x^2-n_y^2)\left(\hat{x}'\hat{p}_y'+\hat{y}'\hat{p}_x'\right)\Bigr]\hat{\sigma}_y'\notag\\
&+\frac{2mE_{\mathrm{Z}}}{3\hbar}
\Bigl[\frac{7}{(m\omega_0)^2}(n_x^2-n_y^2)\hat{p}_x'\hat{p}_y'+
2(n_x^2-n_y^2)\hat{x}'\hat{y}'\notag\\
&+\frac{7n_xn_y}{2(m\omega_0)^2}(3\hat{p}_x'^2-\hat{p}_y'^2)
+n_xn_y(3\hat{x}'^2-\hat{y}'^2)\Bigr]\hat{\sigma}_z\Bigr).
\end{align}

We then rewrite the coordinate/momentum operators into 
Bosonic operators $(\hat{a}, \hat{a}^\dagger)$ for $x'$-direction
and $(\hat{b}, \hat{b}^\dagger)$ for $y'$-direction:
\begin{align}
\hat{x}'=\frac{\ell_0}{\sqrt{2}}(\hat{a}+\hat{a}^\dagger),\ & \
\hat{p}_x'=\frac{\hbar}{i}\frac{1}{\sqrt{2}\ell_0}(\hat{a}-\hat{a}^\dagger),\notag\\
\hat{y}'=\frac{\ell_0}{\sqrt{2}}(\hat{b}+\hat{b}^\dagger),\ & \
\hat{p}_y'=\frac{\hbar}{i}\frac{1}{\sqrt{2}\ell_0}(\hat{b}-\hat{b}^\dagger),
\end{align}
where $\ell_0=\sqrt{\hbar/(m\omega_0)}$.
Since there is no drive for $y'$-direction, we can approximate 
the operators including $\hat{b}$ and $\hat{b}^\dagger$ by their 
expectation values with vacuum.
Then the terms include $\hat{\sigma}_x'$ and $\hat{\sigma}_y'$ disappear.
The final expression of the effective Hamiltonian is
\begin{align}\label{eq:cubicDresselhaus}
\left[\hat{S}_{\mathrm{C}}, \hat{\cal H}_{\mathrm{F}}(t)\right]&=
eE(t)m\hbar\gamma\frac{E_{\mathrm{Z}}}{\hbar\omega_0}n_xn_y\notag\\
\times&\left[ 
1+3\hat{a}^\dagger\hat{a}-\frac{5}{6}(\hat{a}^2+\hat{a}^{\dagger 2})
\right]\hat{\sigma}_z\notag\\
&\equiv -\hat{\cal E}_{\mathrm{C}}(t)\hat{\sigma}_z.
\end{align}

\section{Dynamics of spin and orbital coupled system}\label{sec:dynamics}
\subsection{Driven harmonic oscillator with relaxation}
In this subsection, we first discuss the dynamics of driven harmonic oscillator and
then the effect of the relaxation by coupling to the environment.
The driven orbital Hamiltonian is
\begin{align}
\hat{\cal H}_{\mathrm{orb}}(t)&=\hat{\cal H}_{\mathrm{d2D}}+\hat{\cal H}_{\mathrm{F}}(t).
\end{align}
With rotating the coordinate and focusing on the one-dimensional system in the $x'$-component,
\begin{align}
\hat{\cal H}_{\mathrm{orb}}(t)&=\hbar\omega_0\hat{a}^\dagger\hat{a}
+eE_0 \cos(\omega t)\frac{\ell_0}{\sqrt{2}}(\hat{a}+\hat{a}^\dagger)\notag\\
&\sim \hbar\omega_0\hat{a}^\dagger\hat{a}
+\Omega(\hat{a}e^{i\omega t}+\hat{a}^\dagger e^{-i\omega t}),
\end{align}
where we applied rotating-wave approximation and
defined a frequency 
\begin{align}
\Omega&=\frac{eE_0\ell_0}{2\sqrt{2}\hbar}.
\end{align}

Assuming the orbital system is weakly and linearly coupled to the environment (phonons or photons), 
Gorini-Kossakowski-Sudarshan-Lindblad (GKSL) quantum master equation is 
\begin{align}
\partial_t\hat{\rho}_{\mathrm{orb}}&=
-\frac{i}{\hbar}\left[\hat{\cal H}_{\mathrm{orb}}(t),\hat{\rho}_{\mathrm{orb}}\right]
+{\cal L}\hat{\rho}_{\mathrm{orb}},
\end{align}
where the Liouville superoperator (dissipator) ${\cal D}$ applying to the density operator $\hat{\rho}$
is defined by
\begin{align}
{\cal D}\hat{\rho}&\equiv 
\Gamma(\bar{n}+1)\left(\hat{a}\hat{\rho}\hat{a}^\dagger
-\frac{1}{2}\left\{\hat{a}^\dagger\hat{a}, \hat{\rho}\right\}\right)\notag\\
&+\Gamma\bar{n}\left(\hat{a}^\dagger\hat{\rho}\hat{a}
-\frac{1}{2}\left\{\hat{a}\hat{a}^\dagger, \hat{\rho}\right\}\right).
\end{align}
We have introduced the relaxation rate $\Gamma\ (\ge 0)$
and bath Bosonic distribution function $\bar{n}\equiv 1/(e^{\beta \hbar\omega_0}-1)$
with an inverse temperature $\beta\equiv 1/(k_{\mathrm{B}}T)$.

\subsection{Master equation for the spin and orbital system}
When the spin and orbital degrees of freedom couple, we have the 
total Hamiltonian up-to linear order in SOI,
\begin{align}
\hat{\cal H}_{\mathrm{SW}}(t)&=\hat{\cal H}_{\mathrm{orb}}(t)
+\hat{\cal H}_{\mathrm{Z}}
-\hat{\cal E}(t)\hat{\sigma}_z,
\end{align}
where $\hat{\cal E}(t)\equiv {\cal E}_{\mathrm{L}}(t)+\hat{\cal E}_{\mathrm{C}}(t)$
and $\hat{\cal E}_{\mathrm{C}}$ includes operators $\hat{a}$ and $\hat{a}^\dagger$.
Then the quantum master equation of total system $\hat{W}(t)$ is
\begin{align}
\partial_t\hat{W}&=
-\frac{i}{\hbar}\left[\hat{\cal H}_{\mathrm{SW}}(t),\hat{W}\right]
+{\cal D}\hat{W}.
\end{align}
In general, the quantum master equation should be globally treated when the 
system is made of multiple components\cite{Uchiyama}.
However, since the spin dynamics induced by the SOI and oscillating electric field is slow,
there are separations of the timescale of the spins and orbitals, and
the dissipation described by the Liouville superoperator only contains the operator of 
the orbital part.

\subsection{Rotating frame}
In this subsection, we introduce a unitary operator to provide 
a discussion in the rotating frame which removes explicit time dependence of the Hamiltonian.
We first study unitary evolution ($\Gamma=0$) with the Schr\"{o}dinger equation:
\begin{align}
i\hbar\partial_t \ket{\psi}
&=\hat{\cal H}_{\mathrm{SW}}(t)
\ket{\psi}
\end{align}
With a unitary operator using a real parameters $R$,
\begin{align}
\hat{U}(t)&=e^{i(2R\hat{a}^\dagger \hat{a}+R\hat{\sigma}_x)t},
\end{align}
we define a new wave function
\begin{align}
\ket{\varphi}
&=\hat{U}(t)
\ket{\psi}
\end{align}
Equation of motion of $\left|\varphi\right\rangle$ is
\begin{align}
i\hbar\partial_t
\ket{\varphi}
&=i\hbar\left[\left\{\partial_t\hat{U}(t)\right\}
\ket{\psi}
+\hat{U}(t)\partial_t
\ket{\psi}
\right]\notag\\
&=i\hbar\left\{2iR\hat{a}^\dagger\hat{a}+iR\hat{\sigma}_x\right\}\hat{U}(t)\ket{\psi}
+\hat{U}(t)\hat{\cal H}_{\mathrm{SW}}(t)
\ket{\psi}
\notag\\
&=-\hbar\left(2R\hat{a}^\dagger\hat{a}+R\hat{\sigma}_x\right)\ket{\varphi}
+\hat{U}(t)\hat{\cal H}_{\mathrm{SW}}(t)\hat{U}^{\dagger}(t)\ket{\varphi}.
\end{align}
Now we study the unitary transformation of the operators, 
$\hat{a}^\dagger\hat{a}, \hat{a}, \hat{a}^\dagger$ and $\hat{\sigma}_z$.
Clearly,
\begin{align}
\hat{U}(t)\hat{a}^\dagger\hat{a}\hat{U}^\dagger(t)&=\hat{a}^\dagger\hat{a}.
\end{align}
Next, we define
\begin{align}
\hat{a}(t)&\equiv \hat{U}(t)\hat{a}\hat{U}^\dagger(t),
\end{align}
which satisfies an equation of motion,
\begin{align}
\partial_t\hat{a}(t)&=\hat{U}(t)2iR\left[\hat{a}^\dagger\hat{a},
\hat{a}\right]\hat{U}^\dagger(t)\notag\\
&=2iR \hat{U}(t)(-\hat{a})\hat{U}^\dagger(t)=-2iR\hat{a}(t).
\end{align}
With using the initial condition, $\hat{a}(t=0)=\hat{a}$, we have the solution
\begin{align}
\hat{a}(t)&=\hat{a}e^{-2iR t}.
\end{align}
Similarly, we have
\begin{align}
\hat{a}^\dagger(t)&\equiv \hat{U}(t)\hat{a}^\dagger\hat{U}^\dagger(t)
=\hat{a}^\dagger e^{2iR t}.
\end{align}
We define eigenstates of $\hat{\sigma}_x$ as $\hat{\sigma}_x\ket{\pm}=\pm\ket{\pm}$,
which provide definitions of other operators 
$\hat{\sigma}_+=\ket{+}\bra{-}$, $\hat{\sigma}_-=\ket{-}\bra{+}$
and $\hat{\sigma}_z=\hat{\sigma}_++\hat{\sigma}_-$.
Then, we discuss the evolution of
\begin{align}
\hat{\sigma}_+(t)&\equiv e^{iR\hat{\sigma}_x t}\hat{\sigma}_+e^{-iR\hat{\sigma}_x t},
\end{align}
with an initial condition, $\hat{\sigma}_+(0)=\hat{\sigma}_+$.
This satisfies an equation of motion
\begin{align}
\partial_t\hat{\sigma}_+(t)&=
e^{iR\hat{\sigma}_x t}iR\left[\hat{\sigma}_x, \hat{\sigma}_+\right]
e^{-iR\hat{\sigma}_x t}\notag\\
&=2iR\hat{\sigma}_+(t),
\end{align}
hence we have the solution
\begin{align}
\hat{\sigma}_+(t)&=e^{2iR t}\hat{\sigma}_+,
\end{align}
and by taking Hermite conjugate of this, we have
\begin{align}
\hat{\sigma}_-(t)&=e^{-2iR t}\hat{\sigma}_-.
\end{align}

Therefore, 
\begin{align}
\hat{U}(t)\hat{\cal H}_{\mathrm{SW}}(t)\hat{U}^\dagger(t)&=\hbar\omega_0\hat{a}^\dagger\hat{a}
+\frac{1}{2}E_{\mathrm{Z}}\hat{\sigma}_x\notag\\
&+\hbar\Omega\left(\hat{a}e^{-2iR t+i\omega t}
+\hat{a}^\dagger e^{2iRt-i\omega t}\right)\notag\\
&-\frac{1}{2}\hat{M}\left(\hat{a}_- e^{-2iRt+i\omega t}+\hat{a}_+e^{2iRt-i\omega t}\right),
\end{align}
where we noted $\hat{\cal E}(t)=\frac{1}{2}(e^{i\omega t}+e^{-i\omega t})\hat{M}\hat{\sigma}_z$
where $\hat{M}$ contains operators $\hat{a}^\dagger$ and $\hat{a}$.
Hence we choose $R=\omega/2$ to remove the time-dependence 
from the Hamiltonian and then
\begin{align}
\hat{U}(t)\hat{\cal H}_{\mathrm{SW}}(t)\hat{U}^\dagger(t)
&=\hbar\omega_0\hat{a}^\dagger\hat{a}+\frac{1}{2}E_{\mathrm{Z}}\hat{\sigma}_x\notag\\
&+\hbar\Omega\left(\hat{a}+\hat{a}^\dagger \right)
-\frac{1}{2}\hat{M}\hat{\sigma}_z.
\end{align}
Now the equation of motion of $\left|\varphi\right\rangle$ is
\begin{align}
i\hbar\partial_t
\ket{\varphi}
=&\Bigl\{\hbar(\omega_0-\omega)\hat{a}^\dagger\hat{a}
+\frac{1}{2}\left(E_{\mathrm{Z}}-\hbar\omega\right)\hat{\sigma}_x\notag\\
&+\hbar\Omega(\hat{a}+\hat{a}^\dagger)-\frac{1}{2}\hat{M}\hat{\sigma}_z\Bigr\}
\ket{\varphi},
\end{align}
then we define a reduced Hamiltonian in the rotating frame
\begin{align}
\hat{\cal H}_r&\equiv \hbar \omega_0'\hat{a}^\dagger\hat{a}
+\frac{1}{2}E_{\mathrm{Z}}'\hat{\sigma}_x
+\hbar\Omega(\hat{a}+\hat{a}^\dagger)-\frac{1}{2}\hat{M}\hat{\sigma}_z,
\end{align}
 with defining $\omega_0'\equiv \omega_0-\omega$ and 
 $E_{\mathrm{Z}}'=E_{\mathrm{Z}}-\hbar\omega$.

We then go back to the case with finite dumping $\Gamma>0$ and 
we define unitary transformed density operator in the 
rotating frame
\begin{align}
\tilde{W}(t)&\equiv \hat{U}(t)\hat{W}\hat{U}^\dagger(t),
\end{align}
which obeys modified QME
\begin{align}\label{eq:rqme}
\partial_t\tilde{W}&=-\frac{i}{\hbar}\left[\hat{\cal H}_r,
\tilde{W}\right]+{\cal D}\tilde{W}.
\end{align}
This can be understood since 
\begin{align}
&\partial_t \tilde{W}(t)=\partial_t\hat{U}(t)\hat{W}\hat{U}^\dagger(t)
+\hat{U}(t)\partial_t\hat{W}\hat{U}^\dagger(t)
+\hat{U}(t)\hat{W}\partial_t\hat{U}^\dagger(t)\notag\\
&=i\left\{\omega \hat{a}^\dagger\hat{a}+\frac{\omega}{2}\hat{\sigma}_x\right\}
\hat{U}(t)\hat{W}\hat{U}^\dagger(t)\notag\\
&+\hat{U}\left\{-\frac{i}{\hbar}\left[\hat{\cal H}(t), \hat{W}\right]
+{\cal D}\hat{W}\right\}\hat{U}^\dagger(t)\notag\\
&\ +\hat{U}(t)\hat{W}\hat{U}^\dagger(t)
\left\{-i\omega\hat{a}^\dagger\hat{a}-i\frac{\omega}{2}\hat{\sigma}_x\right\}\notag\\
&=-\frac{i}{\hbar}\left[\hbar\omega_0'\hat{a}^\dagger\hat{a}+
\frac{1}{2}E_{\mathrm{Z}}'\hat{\sigma}_x+
\hbar\Omega\left(\hat{a}+\hat{a}^\dagger\right)-\frac{1}{2}\hat{M}\hat{\sigma}_z, 
\tilde{W}(t)\right]\notag\\
&\ +\frac{\Gamma}{2}(\bar{n}+1)\left(2\hat{a}\tilde{W}(t)\hat{a}^\dagger
-\hat{a}^\dagger\hat{a}\tilde{W}(t)-\tilde{W}(t)\hat{a}^\dagger\hat{a}\right)\notag\\
&\ +\frac{\Gamma}{2}\bar{n}\left(2\hat{a}^\dagger\tilde{W}(t)\hat{a}
-\hat{a}\hat{a}^\dagger\tilde{W}(t)-\tilde{W}(t)\hat{a}\hat{a}^\dagger\right)\notag\\
&=-\frac{i}{\hbar}\left[\hat{\cal H}_r, \tilde{W}(t)\right]
+{\cal D}\tilde{W}(t).
\end{align}

\subsection{Shifted Bosonic operators}
We then introduce shifted creation/annihilation operators to 
make the GKSL master equation to a regular form.
Using a complex number $q$ to be determined later,
\begin{align}\label{eq:regular}
\tilde{a}&=\hat{a}-q,\\\label{eq:regular2}
\tilde{a}^\dagger&=\hat{a}^\dagger-q^*,
\end{align}
which satisfy Bosonic commutation relation.
Then the dissipator proportional to $\bar{n}+1$ is
\begin{align}
&\hat{a}\tilde{W}\hat{a}^\dagger
-\frac{1}{2}\left\{\hat{a}^\dagger\hat{a},\tilde{W}\right\}\notag\\
&=
(\tilde{a}+q)\tilde{W}(\tilde{a}^\dagger+q^*)
-\frac{1}{2}\left\{(\tilde{a}^\dagger+q^*)(\tilde{a}+q),\tilde{W}\right\}\notag\\
&=\tilde{a}\tilde{W}\tilde{a}^\dagger-\frac{1}{2}\left\{\tilde{a}^\dagger\tilde{a},\tilde{W}\right\}
+\frac{1}{2}\left[q^*\tilde{a}-q\tilde{a}^\dagger,\tilde{W}\right].
\end{align}
Similarly, the dissipator proportional to $\bar{n}$ is
\begin{align}
&\hat{a}^\dagger\tilde{W}\hat{a}
-\frac{1}{2}\left\{\hat{a}\hat{a}^\dagger,\tilde{W}\right\}\notag\\
&=
(\tilde{a}^\dagger+q^*)\tilde{W}(\tilde{a}+q)
-\frac{1}{2}\left\{(\tilde{a}+q)(\tilde{a}^\dagger+q^*),\tilde{W}\right\}\notag\\
&=\tilde{a}^\dagger\tilde{W}\tilde{a}-\frac{1}{2}\left\{\tilde{a}\tilde{a}^\dagger,\tilde{W}\right\}
-\frac{1}{2}\left[q^*\tilde{a}-q\tilde{a}^\dagger,\tilde{W}\right].
\end{align}
Hence, 
\begin{align}\label{eq:modL}
{\cal D}\tilde{W}&={\cal D}_0\tilde{W}
+\frac{\Gamma}{2}\left[q^*\tilde{a}-q\tilde{a}^\dagger,\tilde{W}\right],
\end{align}
where ${\cal D}_0$ is the superoperator which replaces all jump operators in ${\cal D}$ 
by those with tildes.
The coherent dynamics term is
\begin{align}
&-i\left[\omega_0'\hat{a}^\dagger\hat{a}+\Omega\hat{a}+\Omega\hat{a}^\dagger
+\frac{1}{2}E_{\mathrm{Z}}'\hat{\sigma}_x-\frac{1}{2}\hat{M}\hat{\sigma}_z,
\tilde{W}\right]\notag\\
&=-i\left[\omega_0'\tilde{a}^\dagger\tilde{a}
+\frac{1}{2}E_{\mathrm{Z}}'\hat{\sigma}_x
-\frac{1}{2}\tilde{M}\hat{\sigma}_z,\tilde{W}\right]\notag\\
&-i\left[(\omega_0'q^*+\Omega)\tilde{a}
+(\omega_0'q+\Omega)\tilde{a}^\dagger,\tilde{W}\right].\label{eq:regular3}
\end{align}
where $\tilde{M}$ is obtained by substitution of Eqs.~(\ref{eq:regular}, \ref{eq:regular2}).
Combining the second term of Eq.~(\ref{eq:modL}) and the second term of Eq.~(\ref{eq:regular3}),
\begin{align}
&\frac{\Gamma}{2}\left[q^*\tilde{a}-q\tilde{a}^\dagger,\tilde{W}\right]
-i\left[(\omega_0'q^*+\Omega)\tilde{a}
+(\omega_0'q+\Omega)\tilde{a}^\dagger,\tilde{W}\right]\notag\\
&=-i\left[\left(\omega_0'q^*+\Omega-\left(\frac{i\Gamma q}{2}\right)^*\right)\tilde{a}
+\mbox{h.c.},
\tilde{W}\right].
\end{align}
Therefore, by choosing $q$ satisfying
\begin{align}
q&=\frac{\Omega}{-\omega_0'+\frac{i\Gamma}{2}},
\end{align}
this extra term can be dismissed and we can reach a generic GKSL master equation
\begin{align}
\partial_t\tilde{W}&=-\frac{i}{\hbar}\left[\hbar\omega_0'\tilde{a}^\dagger\tilde{a}
+\frac{1}{2}E_{\mathrm{Z}}'\hat{\sigma}_x-\frac{1}{2}\tilde{M}\hat{\sigma}_z,
\tilde{W}\right]
+{\cal D}_0\tilde{W}.
\end{align}
%

\subsection{Interaction Hamiltonian}
Now, we discuss the effective Hamiltonian after the regularization
of the Bosonic operators.
Because of the following relation
\begin{align}
\hat{a}^\dagger\hat{a}&=\tilde{a}^\dagger\tilde{a}+\left|q\right|^2
+q\tilde{a}^\dagger+q^*\tilde{a},
\end{align}
the operator $\tilde{M}$ can be expanded as
\begin{align}
\tilde{M}&=\hbar\Omega_{\mathrm{R}}(\Omega)+M_1\left\{\tilde{a}^\dagger\tilde{a}
+q\tilde{a}^\dagger+q^*\tilde{a}\right\},
\end{align}
 where we had neglected the terms $\hat{a}^2$ and $\hat{a}^{\dagger 2}$
 assuming the excitation to the higher levels are not significant.
$\Omega_{\mathrm{R}}(\Omega)$ is the effective Rabi frequency explicitly given by
 \begin{align}
 \Omega_{\mathrm{R}}(\Omega)
 &=\frac{\hbar E_{\mathrm{Z}}eE_0}{(\hbar\omega_0)^2-E_{\mathrm{Z}}^2}
 \left\{\alpha+2n_xn_y\beta\right\}\notag\\
 &-\frac{\hbar E_{\mathrm{Z}} eE_0}{\hbar\omega_0}m\gamma n_xn_y
 \left(1+\frac{3\Omega^2}{(\omega_0')^2+\left(\frac{\Gamma}{2}\right)^2}\right)\notag\\
 &=\Omega_{\mathrm{R}}^{(1)}-c\Omega^3,
 \end{align}
where $\Omega_{\mathrm{R}}^{(1)}$ is linear in the microwave amplitude $E_0\ (\Omega)$ 
 and $c$ is a positive constant.
 We call $\Omega_{\mathrm{R}}(\Omega)$ as Rabi frequency since
if we can neglect $\hat{H}_I$ and at resonant requency, $E_{\mathrm{Z}}'=0$
 ($\omega=E_{\mathrm{Z}}/\hbar$), the spin Hamiltonian is
 \begin{align}
 \hat{\cal H}_{\mathrm{spin}}&=-\frac{1}{2}\hbar\Omega_{\mathrm{R}}(\Omega)\hat{\sigma}_z.
 \end{align}
 and its general solution is
 \begin{align}
 \ket{\psi(t)}&=
 \left[\cos\left(\frac{\Omega_{\mathrm{R}}t}{2}\right)
 +i\hat{\sigma}_z\sin\left(\frac{\Omega_{\mathrm{R}}t}{2}\right)\right]
 \ket{\psi(0)}.
 \end{align}
 When the microwave amplitude small, the dynamics is dominated by the linear SOI
 (except for a singular situation such that $\alpha+2n_xn_y\beta=0$), 
and  the Rabi frequency linearly grows with microwave amplitude $E_0\ (\Omega)$
but eventually {\it saturates} for large amplitude condition (for $n_xn_y>0$ and
$\alpha+2n_xn_y\beta>0$).
In a particular situation that $\alpha+2n_xn_y\beta=0$, 
 $\Omega_{\mathrm{R}}^{(1)}$ is negative and the Rabi frequency becomes super-linear.

Then the final form of the effective Hamiltonian is
\begin{align}
\tilde{\cal H}_r&=\hbar\hbar\omega_0'\tilde{a}^\dagger\tilde{a}
+\frac{1}{2}E_{\mathrm{Z}}'\hat{\sigma}_x
-\frac{1}{2}\hbar\Omega_{\mathrm{R}}(\Omega)\hat{\sigma}_z
+\hat{\cal H}_I,
\end{align}
where we introduced residual spin-orbit interaction term
\begin{align}\label{eq:ressoi}
\hat{\cal H}_I&\equiv -\frac{1}{2}M_1\left\{\tilde{a}^{\dagger}\tilde{a}
+q\tilde{a}^\dagger+q^*\tilde{a}\right\}\hat{\sigma}_z,
\end{align}
where 
\begin{align}
M_1&\equiv \frac{\hbar E_{\mathrm{Z}} eE_0}{\hbar\omega_0}3m\gamma n_xn_y, 
\end{align}
 
 \subsection{Relaxation of spin}
 The effect of residual spin-orbit interaction ($\hat{\cal H}_I$) is investigated for general situation
 in the Appendix~\ref{app1} treating the orbital degree of freedom approximated as a pseudo spin
 assuming the excitation to the third level is not significant.
We show that $\hat{\cal H}_I$ induces relaxation/decoherence of the spin dynamics with using the 
 adiabatic elimination approximation.\cite{Schlosshauer}
 
 First we map the Bosonic operators to pseudo-spin operators like
 $\hat{a}^\dagger+\hat{a}\leftrightarrow \hat{S}_z$, 
 $-i\hat{a}^\dagger+i\hat{a}\leftrightarrow \hat{S}_y$
 and $\hat{a}^\dagger\hat{a}\leftrightarrow \frac{1}{2}\left(\hat{S}_x+\hat{\cal I}\right)$
 where $\hat{\cal I}$ is the pseudo-spin identity operator.
 We identify the pseudo-spin excitation frequency $\Delta=\omega_0$.
 Then the interaction Hamiltonian, Eq.~(\ref{eq:ressoi}), becomes
 \begin{align}
 \hat{\cal H}_I&\to -\frac{M_1}{4}\left\{\hat{\cal I}+\hat{S}_x+(q+q^*)\hat{S}_z+i(q-q^*)\hat{S}_y\right\}
 \hat{\sigma}_z,
 \end{align}
 By comparing the interaction Hamiltonian in Eq.~(\ref{eq:hig}),
 $\hat{A}_x=-\frac{M_1}{4g}\hat{\sigma}_z$, $\hat{A}_y=-\frac{iM_1}{4g}(q-q^*)\hat{\sigma}_z$
 and $\hat{A}_z=-\frac{M_1}{4g}(q+q^*)\hat{\sigma}_z$.
 Then the (Lamb-shift like) renormalization term for the spin Hamiltonian becomes
 \begin{align}
 \hat{\cal H}_{\mathrm{S}}'/\hbar&=-\frac{g}{2\bar{n}+1}\hat{A}_x-\frac{M_1}{4}\hat{\sigma}_z
 =-\frac{2\bar{n}}{2\bar{n}+1}M_1\hat{\sigma}_z\equiv 
 -\frac{1}{2}\Omega_{\mathrm{R}}'\hat{\sigma}_z.
 \end{align}
Then the quantum master equation of the spin at resonant condition is
\begin{align}\label{eq:qmespin}
\frac{d\hat{\rho}_{\mathrm{S}}}{dt}
&=-i\left[-\frac{1}{2}\tilde{\Omega}_{\mathrm{R}}(\Omega)\hat{\sigma}_z,
\hat{\rho}_{\mathrm{S}}\right]
+\frac{1}{2}R_{\mathrm{total}}{\cal D}[\hat{\sigma}_z]\hat{\rho}_{\mathrm{S}}.
\end{align}
We define renormalized Rabi frequency $\tilde{\Omega}_{\mathrm{R}}(\Omega)
\equiv \Omega_{\mathrm{R}}(\Omega)+\Omega_{\mathrm{R}}'$.
There are finite relaxation rate $R_0$ independent of microwave frequency\cite{Tokura},
which is originated from the spin state mixing by SOI and the electron-phonon couplings. 
Hence, total relaxation rate $R_{\mathrm{total}}$ is given by
\begin{align}
R_{\mathrm{total}}(\Omega)&=R_0+R(\Omega),
\end{align}
where microwave frequency dependent 
relaxation rate $R(\Omega)$ is evaluated from Eq.~(\ref{eq:diss})
\begin{align}
R(\Omega)&\equiv \left(\frac{\hbar E_z eE_0}{\hbar\omega_0}3m\gamma n_x n_y\right)^2
\frac{1}{4(2\bar{n}+1)\Gamma}\notag\\
&\times \left\{1+\frac{8(2\bar{n}+1)\Gamma^2}{(2\bar{n}+1)\Gamma^2+4\Delta^2}
\frac{\Omega^2}{(\omega_0')^2+\left(\frac{\Gamma}{2}\right)^2}\right\}.
\end{align}
This rate increases with $\Omega^2$ for small $\Omega$ and additional correction of $\Omega^4$
for larger $\Omega$.

 The general solution of the quantum master equation, Eq.~(\ref{eq:qmespin}), 
 can be easily obtained and the population 
 of the spin in $\ket{+}$ state (positive eigen state of $\hat{\sigma}_x$) $P(t)$ starting from 
 $\ket{+}$ at $t=0$ is
 \begin{align}
 P(t)&=\braket{+| \hat{\rho}(t)|+}\notag\\
 &=\frac{1}{2}\left(1+e^{-R_{\mathrm{total}}(\Omega) t}\cos(\tilde{\Omega}_{\mathrm{R}}(\Omega)t)\right).
 \end{align}
 Therefore, $1/R_{\mathrm{total}}(\Omega)$ is the spin relaxation time.
 
 The ratio $\tilde{\Omega}_{\mathrm{R}}(\Omega)/R_{\mathrm{total}}(\Omega)$ quantifies the
 fidelity of the EDSR.
 In general, this ratio linearly increases with $\Omega$ for small $\Omega$ and
 eventually saturates or decreases for larger $\Omega$.
 
 \section{Ge hole Rashba SOI EDSR}\label{sec:ge}
 The holes in the valence band generally have larger SOI.
 In semiconductors with inversion symmetry like Ge, Dresselhaus SOI is absent.
 However, it had shown that cubic Rashba SOI is
 expected in Ge quantum well and provides possibility of fast EDSR.\cite{Terrazos}
 The SOI Hamiltonian reads
 \begin{align}
 \hat{\cal H}_{\mathrm{R}}&=i\alpha\left(\hat{\pi}_+\hat{\pi}_-\hat{\pi}_+\hat{\sigma}_+
 -\hat{\pi}_-\hat{\pi}_+\hat{\pi}_-\hat{\sigma}_-\right),
 \end{align}
 where $\hat{\pi}_{\pm}\equiv \hat{\pi}_x\pm i\hat{\pi}_y$ and
 $\hat{\sigma}_{\pm}\equiv \hat{\sigma}_+\pm i\hat{\sigma}_y$.
The magnetic field is applied normal to the two-dimensional plane, $\bm{n}=\hat{z}$.
The oscillating electric field is applied in-plane, $\bm{n}'=(n_x, n_y, 0)$ with $n_x^2+n_y^2=1$.
Since the out-of-plane g-factor is so large, the necessary
magnetic field for the EDSR could be small and here we use the approximation
$\hat{\pi}_{\pm}\sim \hat{p}_{\pm}$.
With similar treatment as in Sec.\ref{sec:cubic}, we found the effective driving
Hamiltonian for the spin is
\begin{align}
-{\cal L}_{\mathrm{F}}S(t)&=eE(t)\frac{2\alpha E_{\mathrm{Z}}m}{\omega_0}
\left(1+3\hat{a}^\dagger\hat{a}\right)\hat{\sigma}_y',
\end{align}
where we have rotated the coordinate such that the $x$-axis along to $\bm{n}'$.
This expression is similar to that of cubic Dresselhaus, Eq.~(\ref{eq:cubicDresselhaus}),
except that this Hamiltonian is isotropic.
With similar arguments as before, we can show that the Rabi frequency 
becomes super-linear for larger amplitude of the microwave 
and at the same time, spin relaxation becomes significant because of the
finite residual spin-orbital couplings.
 
 \section{Conclusions}\label{sec:conclusion}
 We have analyzed the electric dipole spin resonance with linear- and cubic-
 spin-orbit interaction (SOI) up-to a linear order of the SOI using Schrieffer-Wolff
 transformation and GKSL-type quantum master equation approach.
 When the amplitude of the microwave becomes larger, the Rabi frequency 
 shows non-linear dependence on the amplitude. 
 Moreover, the residual couplings of spin degree of freedom to the orbital degree of
 freedom induces the relaxation or decoherence of the spin and degrades
 the fidelity of the EDSR coherent spin manipulation.
 We also notice that the EDSR with Rashba SOI for Ge holes in a quantum well
 is quite similar to the cubic Dresselhaus EDSR and non-linear Rabi frequency and
 spin relaxation for larger microwave amplitude are predicted.

\begin{acknowledgements}
We thank useful discussions with Taichi Furuya.
This work is supported from the JST Moonshot R\&D-MILLENNIA program 
(Grant No. JPMJMS2061) and JSPS KAKENHI (Grant No. 23H05458).
\end{acknowledgements}

\appendix
\section{Effective spin master equation using adiabatic ellimination}\label{app1}
\subsection{Model and setups}
We are discussion a single spin coupled to pseudo-spin, which
is dissipative in a GKSL master equation.
The pseudo-spin is modeled from an orbital harmonic oscillator 
assuming the excitation from the ground state is not significant.
The total Hamiltonian is $\hat{\cal H}=\hat{\cal H}_{\mathrm{S}}+\hat{\cal H}_{\mathrm{PS}}+
\hat{\cal H}_{\mathrm{I}}$, whose terms are given by
\begin{align}
\hat{\cal H}_{\mathrm{S}}&=-\frac{1}{2}\hbar\Omega_{\mathrm{R}} \hat{\sigma}_z,\\
\hat{\cal H}_{\mathrm{PS}}&=\frac{1}{2}\hbar\Delta\hat{S}_x,\\
\hat{\cal H}_{\mathrm{I}}&=\hbar g\sum_{\nu=x,y,z}\hat{A}_\nu\hat{S}_{\nu},\label{eq:hig}
\end{align}
where $\hat{A}_{\nu}$ are Hermite operators made of spin operators $\hat{\sigma}_x,
\hat{\sigma}_y, \hat{\sigma}_z$.
We introduce eigenstates of $\hat{S}_x$, $\ket{\pm}$ satisfying the relation
\begin{align}
\hat{S}_x\ket{\pm}&=\pm \ket{\pm},\\
\hat{S}_y\ket{\pm}&=\mp i\ket{\mp},\\
\hat{S}_z\ket{\pm}&=\ket{\mp}.
\end{align}
We also introduce raising (lowering) operator $\hat{S}^+=\ket{+}\bra{-}$ ($\hat{S}^-=\ket{-}\bra{+}$)
which satisfies $\hat{S}^+\hat{S}^-=\ket{+}\bra{+}, \hat{S}^-\hat{S}^+=\ket{-}\bra{-}$.

Then density matrix of total system is $\hat{W}(t)$, which approximately satisfies
following GKSL master equation,
\begin{align}
\frac{d\hat{W}}{dt}&=-\frac{i}{\hbar}\left[\hat{\cal H}, \hat{W}\right]
+\Gamma(\bar{n}+1){\cal D}[\hat{S}^-]\hat{W}+\Gamma\bar{n}{\cal D}[\hat{S}^+]\hat{W},
\end{align}
where the dissipator is defined with an operator of pseudo-spin system, $\hat{A}$,
\begin{align}
{\cal D}[\hat{A}]\hat{W}&\equiv \hat{A}\hat{W}\hat{A}^\dagger
-\frac{1}{2}\left(\hat{A}^\dagger \hat{A}\hat{W}+\hat{W}\hat{A}^\dagger\hat{A}\right),
\end{align}
and Bosonic distribution function
\begin{align}
\bar{n}&\equiv \frac{1}{e^{\beta \hbar\Delta}-1},
\end{align}
where $\beta$ is the inverse-temperature of the environment and 
$\Gamma\ (>0)$ characterizes the strength of the relaxation.
Main assumption in this work is
\begin{align}
g < \omega_{\mathrm{Z}} \ll \Delta, \Gamma,
\end{align}
which may validate the assumption of local dissipation of pseudo-spin,
although it coupled to the spin system and global approach is necessary in general.\cite{Uchiyama}
The density matrix of the spin system only is given by
\begin{align}
\hat{\rho}_{\mathrm{S}}(t)&=\mbox{Tr}_{\mathrm{PS}}\left[\hat{W}(t)\right].
\end{align}

\subsection{Interaction picture and adiabatic ellimination}
We introduce following unitary transformation to move to the interaction picture
\begin{align}
\tilde{W}(t)&\equiv e^{i\hat{\cal H}_{\mathrm{S}}t/\hbar}\hat{W}e^{-i{\cal H}_{\mathrm{S}}t/\hbar},
\end{align}
then all the operators $\hat{A}_\nu$ becomes $\tilde{A}_{\nu}$ and the master equation
becomes
\begin{align}
\frac{d\tilde{W}}{dt}&=-\frac{i\Delta}{2}\left[\hat{S}_x, \tilde{W}\right]
-ig\sum_{\nu}\left[\tilde{A}_{\nu}\hat{S}_{\nu},\tilde{W}\right]\notag\\
&+
\Gamma(\bar{n}+1){\cal D}[\hat{S}^-]\tilde{W}+\Gamma\bar{n}{\cal D}[\hat{S}^+]\tilde{W},
\end{align}
We then evaluate the matrix elements of $\tilde{W}$,
\begin{align}
\frac{d\tilde{W}_{++}}{dt}&=-ig\left(\left[\tilde{A}_x, \tilde{W}_{++}\right]
+\tilde{B}\tilde{W}_{-+}-\tilde{W}_{+-}\tilde{B}^\dagger\right)\notag\\
&-\Gamma(\bar{n}+1)\tilde{W}_{++}+\Gamma\bar{n}\tilde{W}_{--},\\
\frac{d\tilde{W}_{--}}{dt}&=-ig\left(-\left[\tilde{A}_x, \tilde{W}_{--}\right]
+\tilde{B}^\dagger\tilde{W}_{+-}-\tilde{W}_{-+}\tilde{B}\right)\notag\\
&+\Gamma(\bar{n}+1)\tilde{W}_{++}-\Gamma\bar{n}\tilde{W}_{--},\\
\frac{d\tilde{W}_{+-}}{dt}&=-ig\left(\left\{\tilde{A}_x,\tilde{W}_{+-}\right\}
+\tilde{B}\tilde{W}_{--}-\tilde{W}_{++}\tilde{B}\right)\notag\\
&-\frac{1}{2}\left(\Gamma(2\bar{n}+1)+2i\Delta\right) \tilde{W}_{+-},\label{eq:wpm}
\end{align}
where we introduced $\tilde{B}\equiv \tilde{A}_z+i\tilde{A}_y$ and
$[\ ,\ ]$ and $\{\ ,\ \}$ are commutator and anti-commutator, respectively.
The equation for $\tilde{W}_{-+}$ is obtained by taking the Hermite conjugate of
Eq.~(\ref{eq:wpm}).

We derive the effective master equation of 
$\hat{\rho}_{\mathrm{S}}(t)=\tilde{W}_{++}+\tilde{W}_{--}$
by an approximation, which is called as adiabatic ellimination\cite{Schlosshauer}.
In fact, this approximation is based on the separation of the different timescales of
slow spin dynamics and fast pseudo-spin dynamics.
Neglecting $g\tilde{W}_{+-}$ term on the right-hand-side of Eq.~(\ref{eq:wpm})
and assuming that the steady state is quickly established $d\tilde{W}_{+-}/dt=0$, we have
\begin{align}
\tilde{W}_{+-}&\sim \frac{2ig}{\Gamma(2\bar{n}+1)+2i\Delta}
\left(\tilde{W}_{++}\tilde{B}-\tilde{B}\tilde{W}_{--}\right),\\
\tilde{W}_{-+}&\sim \frac{-2ig}{\Gamma(2\bar{n}+1)-2i\Delta}
\left(\tilde{B}^\dagger\tilde{W}_{++}-\tilde{W}_{--}\tilde{B}^\dagger\right).
\end{align}
Now, we introduce Hermite non-positive operator $\tilde{\eta}_{\mathrm{S}}\equiv 
\tilde{W}_{++}-\tilde{W}_{--}$, and following relations
\begin{align}
\tilde{W}_{++}=\frac{1}{2}\left(\tilde{\rho}_{\mathrm{S}}+\tilde{\eta}_{\mathrm{S}}\right),\ 
& \
\tilde{W}_{--}=\frac{1}{2}\left(\tilde{\rho}_{\mathrm{S}}-\tilde{\eta}_{\mathrm{S}}\right).
\end{align}
Then, we have
\begin{align}
&\frac{d\tilde{\eta}_{\mathrm{S}}}{dt}=-ig\left[\tilde{A}_x, \tilde{\rho}_{\mathrm{S}}\right]\notag\\
&-\frac{g^2}{\Gamma(2\bar{n}+1)-2i\Delta}
\left\{\tilde{B}\tilde{B}^\dagger\tilde{\rho}_{\mathrm{S}}
-\tilde{B}\tilde{\rho}_{\mathrm{S}}\tilde{B}^\dagger
+\tilde{B}^\dagger\tilde{\rho}_{\mathrm{S}}\tilde{B}
-\tilde{\rho}_{\mathrm{S}}\tilde{B}^\dagger\tilde{B}\right\}\notag\\
&-\frac{g^2}{\Gamma(2\bar{n}+1)+2i\Delta}
\left\{\tilde{\rho}_{\mathrm{S}}\tilde{B}\tilde{B}^\dagger
-\tilde{B}\tilde{\rho}_{\mathrm{S}}\tilde{B}^\dagger
+\tilde{B}^\dagger\tilde{\rho}_{\mathrm{S}}\tilde{B}
-\tilde{B}^\dagger\tilde{B}\tilde{\rho}_{\mathrm{S}}\right\}\notag\\
&-\Gamma\tilde{\rho}_{\mathrm{S}}-\Gamma(2\bar{n}+1)\tilde{\eta}_{\mathrm{S}}.
\end{align}
We define following rates
\begin{align}
F&\equiv \frac{1}{2i}\left[\frac{2g^2}{\Gamma(2\bar{n}+1)-2i\Delta}
-\frac{2g^2}{\Gamma(2\bar{n}+1)+2i\Delta}\right]\notag\\
&=\frac{4\Delta g^2}{(2\bar{n}+1)^2\Gamma^2+4\Delta^2},\\
K&\equiv \frac{2g^2}{\Gamma(2\bar{n}+1)-2i\Delta}
+\frac{2g^2}{\Gamma(2\bar{n}+1)+2i\Delta}\notag\\
&=\frac{4\Gamma(2\bar{n}+1)g^2}{(2\bar{n}+1)^2\Gamma^2+4\Delta^2}.
\end{align}
Assuming $d\tilde{\eta}_{\mathrm{S}}/dt=0$, we have
\begin{align}
\tilde{\eta}_{\mathrm{S}}
&=-\frac{ig}{\Gamma(2\bar{n}+1)}\left[\tilde{A}_x, \tilde{\rho}_{\mathrm{S}}\right]
-\frac{1}{2\bar{n}+1}\tilde{\rho}_{\mathrm{S}}\notag\\
&-\frac{iK}{2(2\bar{n}+1)\Gamma}
\left\{\left[\tilde{A}_y, \tilde{A}_z\right],\tilde{\rho}_{\mathrm{S}}\right\}\notag\\
&-\frac{iK}{(2\bar{n}+1)\Gamma}\left(\tilde{A}_z\tilde{\rho}_{\mathrm{S}}\tilde{A}_y
-\tilde{A}_y\tilde{\rho}_{\mathrm{S}}\tilde{A}_z\right)\notag\\
&-\frac{iF}{(2\bar{n}+1)\Gamma}\left[\tilde{A}_z^2
+\tilde{A}_y^2,\tilde{\rho}_{\mathrm{S}}\right]\notag\\
&\sim -\frac{ig}{\Gamma(2\bar{n}+1)}\left[\tilde{A}_x, \tilde{\rho}_{\mathrm{S}}\right]
-\frac{1}{2\bar{n}+1}\tilde{\rho}_{\mathrm{S}},
\end{align}
where we had neglected higher order terms.
Hence, the master equation for the spin system is,
\begin{align}\label{eq:diss}
&\frac{d\tilde{\rho}_{\mathrm{S}}}{dt}\notag\\
&=
-i\left[
\frac{-g}{2\bar{n}+1}\tilde{A}_x
+F\left(\frac{\bar{n}}{2\bar{n}+1}\right)\tilde{B}\tilde{B}^\dagger
-F\left(\frac{\bar{n}+1}{2\bar{n}+1}\right)\tilde{B}^\dagger\tilde{B}, 
\tilde{\rho}_{\mathrm{S}}\right]\notag\\
&+\frac{2g^2}{\Gamma(2\bar{n}+1)}{\cal D}[\tilde{A}_x]\tilde{\rho}_{\mathrm{S}}\notag\\
&+\left(\frac{\bar{n}}{2\bar{n}+1}\right)K{\cal D}[\tilde{B}^\dagger]\tilde{\rho}_{\mathrm{S}}
+\left(\frac{\bar{n}+1}{2\bar{n}+1}\right)K{\cal D}[\tilde{B}]\tilde{\rho}_{\mathrm{S}}\notag\\
&=-\frac{i}{\hbar}\left[\tilde{\cal H}_{\mathrm{S}}', \tilde{\rho}_{\mathrm{S}}\right]
+\sum_{i=1,3} \Gamma_i {\cal D}\left[\tilde{C}_i\right]\tilde{\rho}_{\mathrm{S}},
\end{align}
where the first term provides the renormalization of the system Hamiltonian
and the rests are introducing relaxation or decoherence with the rate $\Gamma_i$.

\end{document}